\documentclass{aa}
\usepackage{graphicx}
\begin{document}
\include{special}
\title{Numerical simulations of expanding supershells in dwarf \\ 
irregular galaxies. I. Application to Holmberg~I}

\author{E. I. Vorobyov\inst{1}, U. Klein\inst{2}, Yu. A. Shchekinov\inst{3},
and J. Ott\inst{4}}

\institute{Institute of Physics, Stachki 194, Rostov-on-Don, Russia, and Isaac
Newton Institute of Chile, Rostov-on-Don Branch;
\email{eduard\_vorobev@mail.ru} 
\and
Radioastronomisches Institut der Universit${\rm\ddot a}$t Bonn, Auf
dem H${\rm\ddot u}$gel 71, D-53121 Bonn, Germany;
\email{uklein@astro.uni-bonn.de}  
\and
Rostov University, Sorge 5, Rostov-on-Don, Russia, and Isaac
Newton Institute of Chile, Rostov-on-Don Branch;
\email{yus@phys.rsu.ru}
\and
CSIRO Australia Telescope National Facility, Cnr Vimiera \& Pembroke
Roads, Marsfield NSW 2122, Australia;
\email{Juergen.Ott@csiro.au}
}

\date{}

\abstract{
Numerical hydrodynamical modelling of supernova-driven shell formation is 
performed with a purpose to reproduce a giant HI ring (diameter 1.7~kpc) 
in the dwarf irregular galaxy Holmberg~I (Ho~I). We find that the contrast in HI 
surface density between the central HI depression and the ring is sensitive 
to the shape of the gravitational potential. This circumstance can be used 
to constrain the total mass (including the dark matter halo) of nearly 
face-on dwarf irregulars. We consider two models of Ho~I, which differ by
 an assumed mass of the dark matter halo $M_{\rm h}$. 
The contrast in HI surface density between the central HI depression and 
the ring, as well as the lack of gas expansion in the central hole, 
are better reproduced by the model with a massive halo of 
$M_{\rm h}=6.0\times 10^9~M_{\odot}$ than by that with a small halo of
$M_{\rm h}=4.0\times 10^8~M_{\odot}$, implying that Ho~I is halo-dominated.
Assuming the halo mass of $6.0\times 10^9~M_{\odot}$, we determine the mechanical energy 
required to form the observed ring equal to $(3.0 \pm 0.5) \times 10^{53}$~ergs, 
equivalent $300\pm 50$ Type~II supernovae. The inclination 
of Ho~I is constrained to $15^\circ-20^\circ$ by comparing the modelled 
HI spectrum and channel maps with those observed. 
}

\authorrunning{E. I. Vorobyov et al.}
\titlerunning{Numerical simulations of Holmberg~I}

\maketitle

\section{Introduction}
The interstellar medium of gas-rich dwarf irregular galaxies (dIrr) is usually
dominated by features described as shells, holes, or rings. The number of such 
structures varies from a few (IC~10, Wilcots \& Miller \cite{Wilcots}) to several 
tens (Holmberg II, Puche et al. \cite{Puche}) and their sizes are within several 
hundred parsecs. However, there are a few dIrr's the HI morphology of which is 
totally dominated by a single ring structure of the size comparable to their 
optical extent (Ott \cite{Ott2}). Among them is Holmberg~I, a member 
of the M81 group of galaxies (distance $\sim$~3.6~Mpc). 
Ott et al. (\cite{Ott}) have found that most of 
its HI content (75\%) is localized within a giant ring of 1.7~kpc diameter.
They suggested that strong stellar winds and supernova (SN) explosions might
be responsible for the peculiar HI morphology in Ho~I.

In the present paper we test this hypothesis via detailed numerical hydrodynamical modelling. Two model galaxies are considered: a fast 
rotating galaxy with a massive halo and a slowly rotating one with a 
small halo. We attempt to determine the amount of SNe required to reproduce 
a ring-like morphology of the distribution of the atomic hydrogen in Ho~I. 
We show that modelling of the HI spectrum, HI channel maps, and the radial 
HI profile may constrain the inclination angle and consequently the dark
matter content of Ho~I.

In Sect.~\ref{origin} the origin of a ring-like HI morphology
in Ho~I is discussed. In Sect.~\ref{model} the numerical
hydrodynamical model for simulating the supernova-driven shell
dynamics is formulated. The results of simulations are presented in
Sect.~\ref{results} and Sect.~\ref{discus}. 
Our conclusions are summarized in Sect.~\ref{concl}.

\section{Origin of the peculiar HI morphology in Ho~I}
\label{origin}
\begin{figure}
  \resizebox{\hsize}{!}{\includegraphics{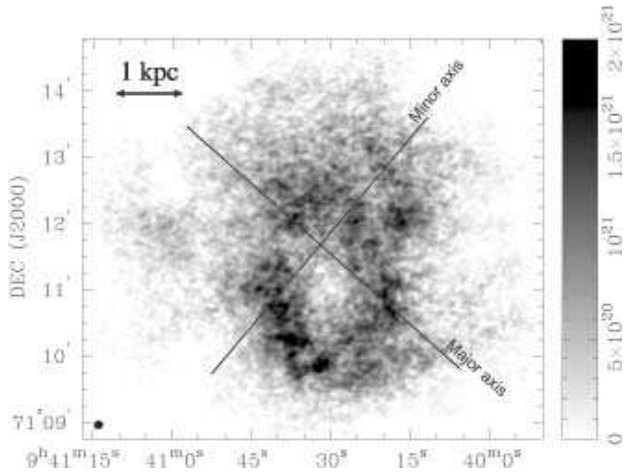}}
      \caption{Integrated HI emission of Ho~I. In the bottom left
      corner we show the half-power beam (8\farcs2 $\times$ 7.0\farcs).
      The grayscale represents the HI column density.
      The data have been obtained from the Very Large Array observations 
      (Ott et al. \cite{Ott}).}
         \label{Fig1}
\end{figure}

Fig.~\ref{Fig1} shows the integrated HI emission of Ho~I with the kinematical 
major and minor axes overlaid (Ott et al. \cite{Ott}). Most of the HI content is 
obviously concentrated in a ring, the morphological center of which is offset 
by 0.7~kpc from the dynamical center of the galaxy. The contrast in column 
density between the central HI depression and the ring is about a factor of 
15. In principle, kpc-sized ring structures in the gas component of disk 
galaxies can be caused by a number of phenomena. Among them are resonant 
rings (Buta \cite{Buta}), collisionally induced rings (Lynds \& Toomre 
\cite{Lynds}), the infall of massive gas clouds (Tenorio-Tagle et al. 
\cite{TT}), gamma ray bursts (Efremov et al. \cite{Efremov}), or multiple 
supernova explosions. In case of Ho~I, the first two scenarios can be ruled 
out because of the absence of spiral structure, the apparent 
absence of lower-mass companions 
along the projected minor axis of Ho~I, and the lack of expansion motions in the 
ring. The latter three scenarios are plausible. However, the HI ring in Ho~I
shows evidence for a supernova-driven origin, since the B-band image of Ho~I indicates
that there is a pronounced concentration of young blue stars within the ring 
(Ott et al. \cite{Ott}), and the most massive stars of this population 
might have been a plausible source 
of energy input over the past 50~Myr in our model. 
In the following sections the supernova-driven mechanism is numerically 
investigated. Other possibilities of HI ring formation in Ho~I will be 
discussed in a subsequent paper.

\section{Numerical model}
\label{model}
\subsection{Initial conditions}
We consider an axisymmetric model galaxy consisting of a rotating gas disk, 
a rigid stellar disk, and a rigid, spherically symmetric halo. Extensive 
radio and optical observations of Ho~I by Ott et al. (\cite{Ott}) provided
important observational constraints on the total mass and density 
distribution of different components of our model galaxy. The HI mass of 
Ho~I is $1.1 \times 10^8~M_{\odot}$, with the total gas mass 
($M_{\rm HI}+M_{\rm He}+M_{\rm H_2})$ amounting to $1.5 
\times 10^8~M_{\odot}$. Using a {\it B-}band luminosity $L_{\rm B}=1.0 
\times 10^8~L_{\odot}$ and a mass-to-light ratio $M_{\rm s}/L_{\rm B}=1$, 
Ott et al. (2001) derived an estimate of the luminous stellar mass, 
$M_{\rm s}=1.0 \times 10^8~M_{\odot}$, in Ho~I. For computational purposes, 
the density profile of the stellar component is chosen as:
\begin{equation}
\rho_{\rm s}=\rho_{\rm s}^0~{\rm sech}(z/z_{\rm s})~\exp(-r/r_{\rm s}),
\label{stdens}
\end{equation}
where $\rho_{s}^{0}$ is the stellar density in the center of the galaxy, 
and $z_{\rm s}$ and $r_{\rm s}$ are the vertical scale height and radial 
scale length of the stellar component, respectively. The radial scale length 
$r_{\rm s}$ can be estimated from the {\it $I_{\rm c}$}-band radial surface 
brightness profile of Ho~I (Ott et al. \cite{Ott}). If the mass follows the
light at the longer wavelengths, it results in $r_{\rm s}\approx 1.5$~kpc.
We have adopted a value of 300~pc for the vertical scale height $z_{\rm s}$, 
which is typical for dwarf irregulars. With $r_{\rm s}$ and $z_{\rm s}$ being 
fixed, the stellar density $\rho_{\rm s}^0$ at $(z,r)=0$ is varied so as to 
obtain the measured luminous stellar mass of $M_{\rm s}=1.0 \times 
10^8~M_{\odot}$ within the computational domain. This results in 
$\rho_{s}^{0}\approx0.02~M_{\odot}$~pc$^{-3}$.

The dark matter content of Ho~I is uncertain. We assume that the density 
profile of the halo can be approximated by a modified isothermal sphere 
(Binney \& Tremaine \cite{Binney})
\begin{equation}
 \rho_{\rm h}= {\rho_{\rm h0} \over 1+(r/r_{\rm h})^2},
\end{equation}
where the central density  $\rho_{\rm h0}$
and  the characteristic scale length $r_{\rm h}$ were given by Mac Low
\& Ferrara (\cite{MF}) and Silich \& Tenorio-Tagle (\cite{Silich}):
\begin{equation}
\rho_{\rm h0}=6.3 \times 10^{10} \left( {M_{\rm h} \over M_{\odot}}
\right)^{-1/3}~h^{-1/3}~M_{\odot}~{\rm kpc}^{-3}
\label{halodens}
\end{equation}
\begin{equation}
 r_{\rm h}=0.89 \times 10^{-5} \left( {M_{\rm h} \over
 M_{\odot}}\right)^{1/2}~h^{1/2}~{\rm kpc}.
\label{halorad}
\end{equation}
Here, $h$ is the Hubble constant in units of 100~km~s$^{-1}$~Mpc$^{-1}$ and 
$M_{\rm h}$ is the total halo mass. We adopt $h=0.65$ throughout the paper.

\begin{figure}
  \resizebox{\hsize}{!}{\includegraphics{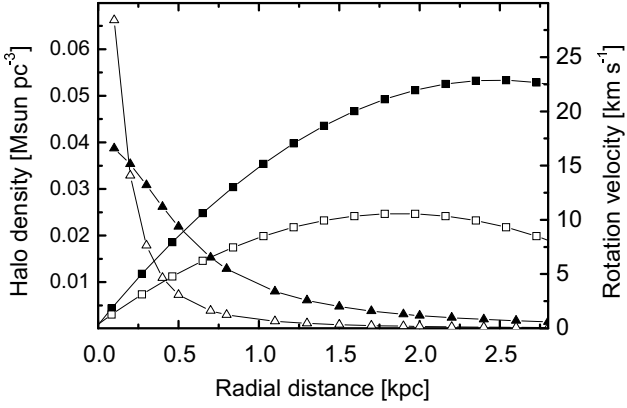}}
      \caption{The radial density profiles of the halo and the initial
      rotation curves for model~1 (filled triangles and filled squares,
      respectively) and model~2 (open triangles and open squares,
      respectively). 
      }
         \label{Fig2}
\end{figure}

We consider two different initial models that attempt to represent the 
appearance of Ho~I before the ring has been formed. \\
a) {\it Fast rotating galaxy with a massive halo} (hereafter model~1). In 
model~1 the gravity force of a massive halo of $6.0 \times 10^9~M_{\odot}$ 
and a stellar disk of $M_{\rm s}=1.0 \times 10^8~M_{\odot}$ is balanced by 
a fast rotating gas disk. We note that the adopted halo mass of 
$6.0\times 10^9~M_{\odot}$ is in fact the total halo mass of Ho~I. 
Particularly, the halo mass confined within the
HI diameter of Ho~I (5.8~kpc) is $3.3 \times 10^8~M_{\odot}$, which 
agrees with the corresponding upper limit of $3.1 \times 10^8~M_{\odot}$
derived by Ott et al. (\cite{Ott}). 
We vary the rotation curve until the initial gas 
surface density distribution resembles that of Ho~I at radii that are 
unperturbed by shell expansion, $r>2$~kpc. We also make sure that the total 
HI mass within the computational area is equal to the measured one ($M_{\rm 
HI} = 1.1 \times 10^8~M_{\odot}$). For the adopted halo mass, the central 
density and characteristic scale length become 
$\rho_{\rm h0}=0.04~M_{\odot}$~pc$^{-3}$ and $r_{\rm h}=0.55$~kpc, 
respectively. The adopted halo density profile and the gas rotation curve 
are shown in Fig.~\ref{Fig2} by the filled triangles and filled squares, 
respectively.  \\
b) {\it Slowly rotating galaxy with a small halo} (hereafter model~2). In 
model~2 the gravity force of a small halo of $M_{\rm h}=4 \times 
10^8~M_{\odot}$ and a stellar disk of $M_{\rm s}=1.0 \times 
10^8~M_{\odot}$ are balanced by a slowly rotating gas disk. For the 
adopted halo mass, the central density and characteristic scale length 
become $\rho_{\rm h0}=0.1~M_{\odot}$~pc$^{-3}$ and $r_{\rm h}=0.14 $~kpc,
respectively. Thus, in a slowly rotating galaxy the halo density has a 
rather cuspy profile as shown by the open triangles in Fig.~\ref{Fig2}.

\subsection{Equilibrium solution}
\label{Steady}
Once the stellar and halo density profiles are fixed, we can proceed to 
obtain the initial gas density distribution by solving the steady-state
momentum equation,
\begin{equation}
{1 \over \rho} {\bf\nabla} P = - ({\bf v} \cdot {\bf\nabla}){\bf v} - 
{\bf\nabla}\Phi_{\rm h} - {\bf\nabla}\Phi_{\rm s},
\label{steady}
\end{equation}
where $P$ is the gas pressure, $\rho$ is the gas density, $\Phi_{\rm h}$ and 
$\Phi_{\rm s}$ are the gravitational potentials of the halo and stellar disk, 
respectively. We assume the gas distribution to be initially symmetric with 
respect to the rotation axis and equatorial plane, which naturally implies the 
use of cylindrical coordinates $(z,r)$ for solving Eq.~(\ref{steady}). If the
gas is initially isothermal, then Eq.~(\ref{steady}) can be transformed to 
the following set of equations:
\begin{eqnarray}
{d~\ln\rho \over dr} &=& {1\over \sigma^2}{v_{\rm rot}^2 \over r} - {1 \over 
\sigma^2} \Big(
{\bf\nabla}\Phi_{\rm h} \Big)_{r} - {1 \over \sigma^2} {d\Phi_{\rm s} \over 
dr} \label{system1}, \\
{d~\ln\rho \over dz} &=& - {1\over \sigma^2} \Big( {\bf\nabla}\Phi_{\rm h} 
\Big)_{z} - {1\over \sigma^2} {d\Phi_{\rm s} \over dz}.
\label{system2}
\end{eqnarray}
Here, $\sigma$ is the gas velocity dispersion. Ott et al. (2001) found an 
average value of $\sigma=9$~km~s$^{-1}$ for Ho~I. The halo gravity force 
can be expressed as:
\begin{eqnarray}
\nabla\Phi_{\rm h}\equiv \nabla\Phi_{\rm h}(r_{*})&=&4 \pi G \rho_{\rm h0}
r_{\rm h}^3/r_{*}^2 \times \nonumber \\
&& \left[ r_{*}/r_{\rm h} - \arctan(r_{*}/r_{\rm h}) \right] {\bf e}_{\rm
r_*},
\end{eqnarray}
where $r_{*}^2=r^2+z^2$ is the radial distance from the galactic center, $r$ 
is the galactocentric radius, $z$ is the height above the midplane, and 
${\bf e}_{\rm r_*}={\bf r_*}/{r_*}$ is a unit vector. The {\it r-} and 
{\it z-}components of the halo gravity force $({\bf\nabla}\Phi_{\rm h})_{\rm 
r}$, $({\bf\nabla}\Phi_{\rm h})_{\rm z}$, can be found by projecting 
$\nabla\Phi_{\rm h}$ on the corresponding coordinate planes. The stellar 
gravitational potential is found by numerically solving the Poisson equation:
\begin{equation}
\triangle\Phi_{\rm s}= 4 \pi G \rho_{\rm s}.
\end{equation}

\begin{figure}
  \resizebox{\hsize}{!}{\includegraphics{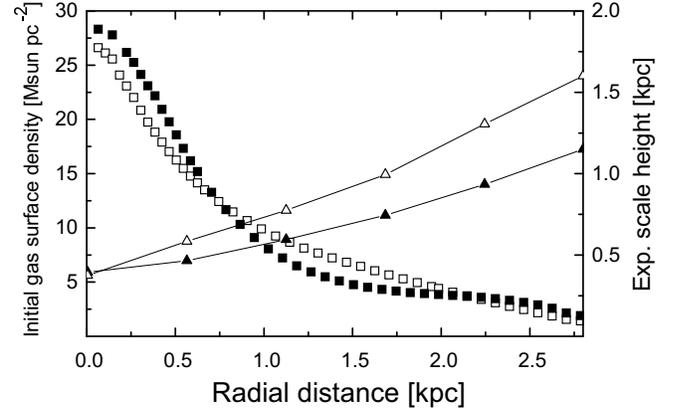}}
      \caption{The equilibrium radial surface density profiles and the vertical 
       scale height of the gas as a function of galactocentric radius for 
       model~1 (filled squares and filled triangles, respectively) and 
       model~2 (open squares and open triangles, respectively).
      }
         \label{Fig3}
\end{figure}

Eqs. (\ref{system1}) and (\ref{system2}) are finite-differenced on a mesh of 
$350 \times 300$ grid points in radial and vertical directions, 
respectively, and solved using the lower-upper decomposition 
method to obtain the equilibrium gas density distribution in a computational 
domain of 3.5~kpc$\times$3.0~kpc. The equilibrium radial surface density 
profiles and the exponential scale heights of the gas for both models are 
shown in Fig.~\ref{Fig3}. In both models the gas has an exponential radial 
profile, which is in agreement with observations of the HI distribution in 
many dIrr's (Taylor et al. \cite{Taylor}). There is also a pronounced 
difference between the models with massive and small halos, the latter has 
a thicker gas disk. Indeed, in model~2 the cuspy halo mainly pulls the matter 
toward the nucleus, while in model~1 the halo density has a somewhat shallower 
distribution, thus contributing significantly to the vertical gravity pull at 
larger radii and subsequently assisting to maintain a thinner equilibrium gas 
configuration.

\subsection{Computational techniques}
We use the ZEUS-2D numerical hydrodynamical code incorporating a rotating
gas disk initially in equilibrium in the fixed stellar and halo potentials.
A usual set of hydrodynamical equations in cylindrical coordinates is solved
using the method of finite-differences with a time-explicit, operator split
solution procedure described in detail in Stone \& Norman (\cite{Stone}).
We have implemented the cooling curve of Spaans \& Norman (\cite{Spaans}) 
and Wada \& Norman (2001)
for a metallicity of one tenth of solar, which is close to the value of 
$Z=Z_{\odot}/12$ derived by Miller \& Hodge (\cite{Miller}) for Ho~I.
The use of cooling function simplifies  the implementation of cooling by 
collecting the effects of various coolants. The cooling processes taken into 
account are: 
(1) recombination of H, He, C, O, N, Si, and Fe; 
(2) collisional excitation of HI, CI-IV, and OI-IV; 
(3) hydrogen and helium bremsstrahlung; 
(4) vibrational and rotational excitation of $H_2$; 
(5) atomic and molecular cooling due to fine-structure emission of C, C+, 
and O and rotational line emission of CO and $H_2$.
We note that using an equilibrium cooling function,
oversimplified particularly in the intermediate-temperature range of $T \le 
10^4$ K, is a common practice in modelling the SN-driven shell dynamics (see e.g. 
Mac Low \& Ferrara \cite{MF}, Murakami \& Babul \cite{MB}). 
Such an oversimplification is justified by 
the fact that most amount of the thermal energy of the shell is lost in 
the high-temperature ($T\sim 10^5-10^6$ K) 
layers, while the energy lost at the low-temperature end contributes less 
to the total energy balance. Variations of the cooling rate connected 
with the fractional ionization are important mostly to determine the cooling time 
of the low- and intermediate-temperature layers. 
However, because at these temperatures 
and the corresponding densities the cooling time is very short in 
comparison to the dynamical time, the overall dynamics is insensitive to details 
of the cooling rate at $T<10^4$ K.
 
We use an empirical heating function tuned to balance the cooling in 
the background atmosphere so that it maintains the gas in hydrostatic 
and thermal equilibrium and may be thought of as a crude model for 
the stellar energy 
input. The disadvantage of this method is that the time-independent heating 
becomes strongly unbalanced in the hot regions filled with supernova ejecta. 
This may result in a spurious energy input at later phases of the shell 
expansion when it occupies a large computational volume. To minimize this 
numerical effect, we have introduced a tracer field function defined by Yabe 
\& Xiao (\cite{Yabe}). The tracer field is advected with the same algorithm 
as the gas density in order to follow the hot gas ejected by the supernovae. 
Heating is prohibited in the regions where the tracer field is present, 
thus largely reducing the effects of spurious heating of the bubble's interior
by the time-independent heating function. This 
is physically justified since most of the heating in the warm interstellar medium 
comes from the photoelectric heating of polycyclic aromatic hydrocarbon molecules 
(PAHs) and small grains, which will be 
either evaporated or highly ionized in the hot ($>10^6$~K) bubble filled with 
supernova ejecta. Cooling and heating are treated numerically using Newton-Raphson iterations, supplemented by a bisection algorithm for occasional zones where the Newton-Raphson method does not converge. In order to monitor accuracy, the total change in the internal energy density in one time step is kept below 
$15\%$. If this condition is not met, the time step is reduced and a solution 
is again sought.

The energy of supernova explosions is released in the form of thermal energy 
in the central region with a radius of four zones. We use a constant wind 
approximation described in detail in Mac Low \& Ferrara (\cite{MF}). 
There is 
a certain degree of freedom in how to distribute $10^{51}$~ergs of mechanical 
energy released by a single supernova. We decided to convert it totally into 
a thermal energy, since in the present simulations we deal with large stellar 
clusters with hundreds of supernovae. With such an amount of SN explosions, 
the surrounding ISM  will be quickly heated and diluted, making radiative 
cooling inefficient. 
We admit that at the initial stages of the stellar cluster evolution 
part of the energy of SNe may be radiated away due to the radiative cooling. 
Hence, our numerical simulations provide a lower bound on 
the number of SN explosions needed to create the observed ring.
We choose the energy input phase to last for 30~Myr, 
which roughly corresponds to a difference in the lifetimes of the most and 
least massive stars capable of producing SNe in a cluster of simultaneously 
born stars.

\section{Multiple Supernova Explosions}
\label{results}
\subsection{Fast rotating galaxy with a massive halo (model~1).}
\label{model1}
\begin{figure*}
 \centering
  \includegraphics{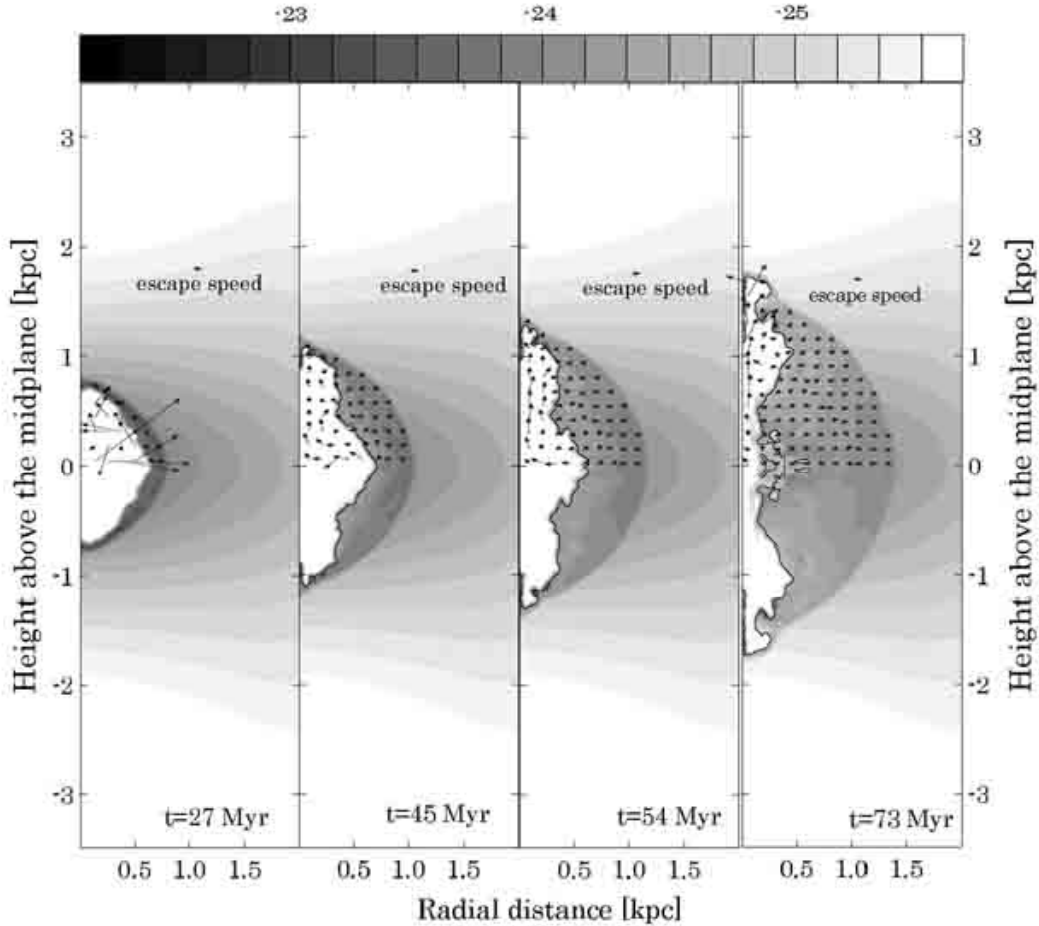}
      \caption{Temporal evolution of the gas volume density distribution 
      in model~1, with the energy input equivalent to 300 successive SN
      explosions. The velocity field is normalized by the {\it local} escape speed.
      In general, the escape speed is a function of both the galactocentric 
      distance and the height above the midplane. Normalizing the velocity field by 
      the value of {\it local} escape speed helps to differentiate the gas
      that can potentially escape the galaxy.
      The contour line delimits the region filled with SN-ejected material. 
      The grey-scale bar is in log units of g~cm$^{-3}$.}
         \label{Fig4}
\end{figure*}
We vary the energy input of the starburst in terms of SN explosions in 
order to reproduce a ring-like gas morphology of Ho~I. We start by showing 
in Fig.~\ref{Fig4} the temporal evolution of the distribution of the gas 
volume density for an energy input equivalent to 300 successive SN 
explosions. SNe generate a supersonically expanding wind that compresses 
the surrounding gas, thus creating a bubble filled with hot ejected gas. 
At the end of the energy input phase ($t=27$~Myr) the shell loses its 
spherical form transforming into a prolate spheroid. As a result, 
tangential motion vortices 
develop in the hot ejected gas, which in turn trigger Kelvin-Helmholtz 
instabilities in the compressed expanding gas layer of the shell. This 
results in a characteristic ripple-like form of the shell. At approximately 
40~Myr the shell breaks out of the disk, Rayleigh-Taylor instability ensues 
from the shell acceleration, creating a characteristic bubble-and-spike 
morphology, first mentioned in numerical experiments by Mac Low et al. 
(\cite{ML}). Soon after the shell stalls and starts to collapse in radial 
direction, simultaneously pushing the hot ejecta to a higher altitude. In 
fact, part of the galaxy ISM will eventually be lost by Ho~I, because its 
velocity exceeds the local escape speed. We compute 
the fraction of the  total gas mass of Ho~I lost by such an outflow
at the time when the ring has collapsed ($t=72$~Myr) by summing up 
the mass of the gas expanding 
with velocities exceeding those of the local escape speed. 
The fraction is low, usually within $1-2\%$.
At $t \approx 54$~Myr the shell appears as a ring if viewed face-on, and
its diameter matches best that of HI ring in Ho~I. At later times of the
evolution ($t \ge 70$~Myr), the central depression is totally filled and 
Ho~I appears as a dwarf irregular, with a declining average gas surface 
density profile. Simulations indicate that for a smaller starburst capable 
of producing only 100 - 150~SNe, the shell never breaks out of the disk and 
hardly expands to 1.7~kpc diameter, indicating that 100-150~SNe are
not sufficient to account for the size of 
HI ring in Ho~I.  For a larger starburst of 450-600~SN, the shell breaks out 
after approximately 35~Myr. However, the radial expansion of the shell 
proceeds to a larger radius than that of Ho~I's ring, mainly because of 
the higher initial momentum acquired by the shell.

The total $\rm H\alpha$ luminosity of Ho~I is $4.3 \times 
10^{38}$~ergs~s$^{-1}$ (Miller \& Hodge \cite{Miller}), which implies a
star formation rate of only $0.004~M_{\odot}$~yr$^{-1}$. We searched for 
dense gas clumps in the expanding shell that could be Jeans unstable. We 
assume that the gas in a restricted computational domain becomes 
Jeans unstable if $t_{\rm ff} < \triangle r /{v_{\rm s}}$, where 
$t_{\rm ff}$ is the free-fall time, $\triangle r$ is the size of a 
computational cell, and $v_{\rm s}$ is the local speed of sound.
We also combined the neighbouring cells to search for a larger scale 
($\le 40 ~pc$) Jeans unstable gas clumps. Our simulations show that 
on the scales of a giant molecular cloud ($\sim 10-40$~pc) the 
shell is stable against self-gravity.
We note however that our simulations were restricted in resolution to 
10~pc, and thus we did not follow possible gravitational instability 
on smaller scales ($<1$~pc) corresponding to the subunits and the cores 
of molecular clouds that can be in principle Jeans unstable, 
while the clouds as a whole are stable (Scoville \& Sanders \cite{SS}).

\begin{figure}
  \resizebox{\hsize}{!}{\includegraphics{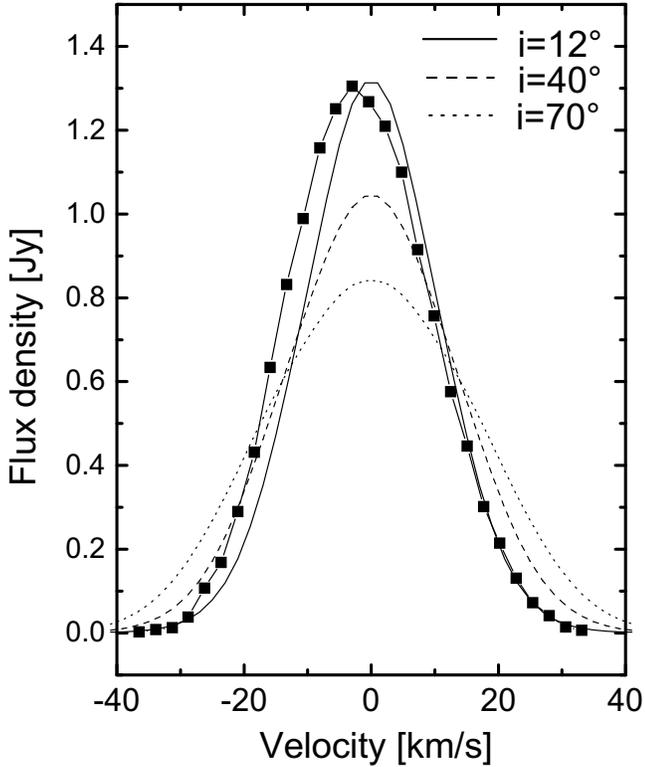}}
      \caption{Model HI spectrum of Ho~I obtained in model~1 for three assumed
      angles of inclination, i = 12$^{\circ}$, 40$^{\circ}$, and 
      70$^{\circ}$. The measured HI spectrum is plotted with the filled 
      squares.}
         \label{Fig5}
\end{figure}

Furthermore, we attempt to reproduce the HI spectrum of Ho~I, shown in 
Fig.~{\ref{Fig5} by the filled squares (Ott et al. \cite{Ott}) and scaled 
to Ho~I's systemic velocity of 141.5~km~s$^{-1}$. Its shape is remarkably 
well described by a Gaussian with a FWHM of $ 27.1$~km~s$^{-1}$. In order 
to model the HI spectrum, we assume that the gas is thermalized in each 
computational cell, with the velocity dispersion controlled by the local 
gas temperature. We also assume that the HI distribution in Ho~I is optically 
thin, which is justified considering its rather low HI surface density, 
$\Sigma_{\rm HI} \le 10~M_{\odot}$~pc$^{-2}$. We use the following 
conversion formula that links the HI mass per velocity channel $M_{\rm HI} 
/\triangle v$ with the HI flux density $S_{\rm HI}$ (Binney \& Merrifield \cite{BM}):
\begin{equation}
{M_{\rm HI} \over \triangle v} = 2.35 \times 10^5~\left( {D \over {\rm Mpc}} 
\right)^{2} \left({S_{\rm HI} \over {\rm Jy}}
\right) \ \   \left[{M_{\odot} \over {\rm km\: s^{-1}}}\right],
\label{HImass}
\end{equation}
where $D$ is the distance to Ho~I.

We have constructed the model HI spectrum of Ho~I for different inclination
angles. We can surprisingly well reproduce the measured HI spectrum for a
small inclination angle, i = 12\degr, represented by the solid line in 
Fig.~\ref{Fig5}; the spectrum for smaller inclinations practically 
coincide with the $i=12^\circ$ one. The model HI spectrum is taken 
at $t=54$~Myr when the 
face-on (or zero-inclination) gas distribution best reproduces that of Ho~I. 
A small difference between the peak positions in the measured and model 
profiles most probably indicates that there is a small offset between the 
morphological center of the ring and the dynamical center of Ho~I, a 
feature supported by the measurements of its HI velocity field (Ott et al. 
\cite{Ott}). Another possibility is that the starburst responsible for 
creating the HI ring is not exactly located in the midplane, but instead 
closer to the near side of the galaxy.

As we employ progressively higher inclination angles, the agreement 
between the model and the measured HI spectrum deteriorates. 
In Fig.~\ref{Fig6} we plot the relative discrepancy between the model and 
measured HI spectrum averaged over 70 velocity channels, starting at 
$-35$~km~s$^{-1}$ and ending at $+35$~km~s$^{-1}$. The flux-weighted 
relative discrepancy is also calculated to de-emphasize the low-mass 
wings of HI spectrum. The best agreement is obviously found for low 
inclination angles, i~$\la$20\degr. The near-Gaussian shape of the HI 
spectrum indicates that the thermal+turbulent motions dominate the
line-of-sight rotation+expansion motions in Ho~I.

\begin{figure}
  \resizebox{\hsize}{!}{\includegraphics{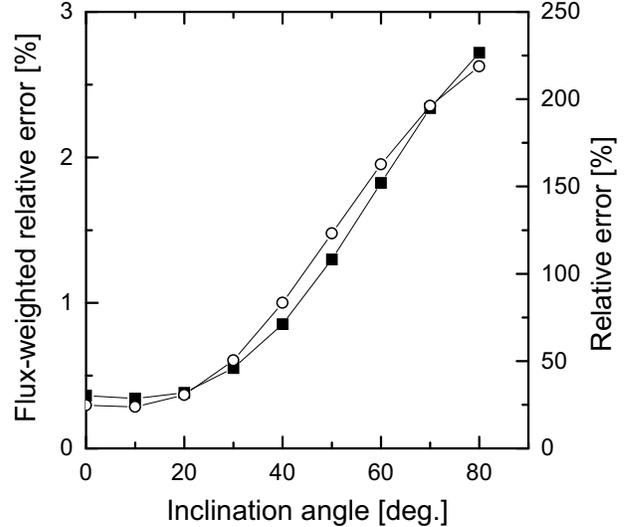}}
      \caption{The relative discrepancy (open circles) between the 
      model and measured HI spectrum of Ho~I as a function of assumed
      inclination obtained for model~1. The flux-weighted relative discrepancy (filled squares) 
      is also provided to de-emphasize the low-mass wings of the HI spectrum.}
         \label{Fig6}
\end{figure}

As is seen from Fig.~\ref{Fig6}, modelling of the HI spectrum provides 
an upper limit to the inclination of Ho~I. A lower limit can be obtained 
by constructing the model HI channel maps and comparing them with the 
observed ones published by Ott et al. (\cite{Ott}). The latter show an 
interesting tendency: the complete ring appears only in a few channel 
maps near the systemic velocity, while the other channel maps show only 
part of it. Specifically, the part of the ring that is faint on the 
approaching side seems to be luminous on the receding side of Ho~I and 
vice versa. We smoothed the [128-131]~km~s$^{-1}$ channel map of Ho~I 
(see Fig.~3 of Ott et al. \cite{Ott}) to a resolution of 20\farcs6~$\times$
21\farcs2 and found that the relative amplitude of the azimuthal variations 
in the HI intensity along the ring is a factor of 2.5. Furthermore, we 
constructed model channel maps at t = 54~Myr for inclinations of 
0\degr~-~20\degr. We find that the observed variations in HI intensity 
along the ring can be obtained only if the inclination is constrained  
to the range 15\degr~-~20\degr. 
As an example, we plot the model channel 
maps constructed at t = 54~Myr for an inclination of 15\degr\, in 
Fig.~\ref{Fig7}. It is readily seen that the area near the central position 
shows virtually no emission throughout all channels, a feature also reported 
by Ott et al. (\cite{Ott}). The asymmetry of the model channel maps indicates 
that the gas is participating in both, rotation and expansion motions, the 
latter dominating at higher altitudes (see Fig.~\ref{Fig4}). The relative
amplitude of azimuthal variations in the maximum HI intensity along the 
ring in the [-13.5; -10.5]~km~s$^{-1}$ channel map, which is equivalent to 
the observed [128-131]~km~s$^{-1}$ channel map of Ott et al. (\cite{Ott}), 
is a factor of 2. This value decreases, as we take progressively lower 
inclination angles and vanishes at zero inclination. Hence, if Ho~I is 
halo dominated as assumed in model~1, then its inclination angle should 
be limited to a range of 15\degr~-~20\degr. 
The lower right panel in 
Fig.~\ref{Fig7} shows the modelled, integrated HI emission. 
Visual inspection shows that the size of the ring
and the HI column density distribution are similar to those in Fig.~\ref{Fig1}, 
indicating that the HI ring can indeed be a result of 
multiple SN explosions. 

The rotation curve of model~1 (the filled squares, Fig.~\ref{Fig2}) 
agrees well with the observed HI rotation curve of Ho~I for $i=15\degr$ 
(see Ott et al. \cite{Ott}, Fig. 9)
only in the outer regions at $r \ga 1.2$~kpc, which are unperturbed by 
the shell expansion.
In the inner perturbed regions the observed velocities exceed 
those of model~1 by 5-10~km~s$^{-1}$. We note here that this discrepancy 
might be due to an offset location of the dynamical center 
of Ho~I with respect to the geometrical center of the HI hole (see Fig.~\ref{Fig1}), 
the effect that can hardly be modelled in our two-dimensional axisymmetric simulations.
As a result, the mass redistribution due to the off-center SN explosions
might have brought a considerable fraction of the fast rotating gas 
towards the galactic center.
Indeed, adopting the observed rotation curve of Ho~I as the initial rotation curve
of model~1
would result in the equilibrium gas distribution that totally lacks any
gas in the central 1~kpc radius irrespective of the halo mass we use;
a realistic equilibrium gas distribution cannot be obtained in this case, 
implying that the observed rotation curve of Ho~I is strongly perturbed 
in the inner regions.

\begin{figure}
  \resizebox{\hsize}{!}{\includegraphics{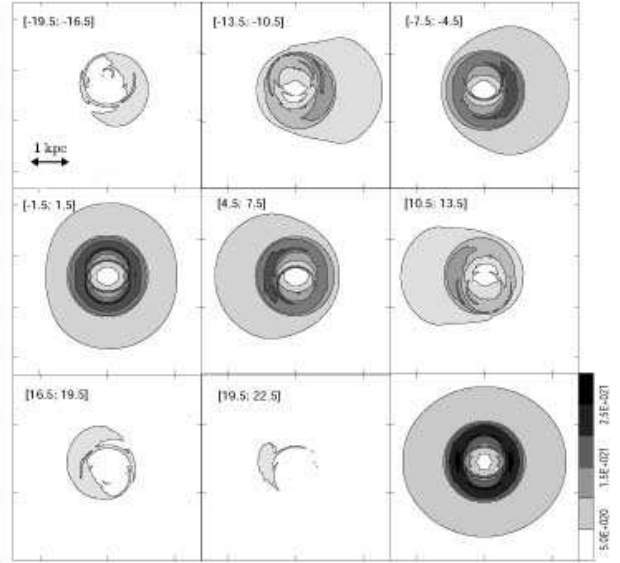}}
      \caption{Model channel maps of Ho~I at the assumed inclination of 
      i = 15\degr. The lower right panel shows the integrated HI emission 
      obtained in model~1. The grey-scale wedge is in units of cm$^{-2}$ 
      and the velocity channels are in units of km~s$^{-1}$. }
         \label{Fig7}
\end{figure}

\begin{figure}
  \resizebox{\hsize}{!}{\includegraphics{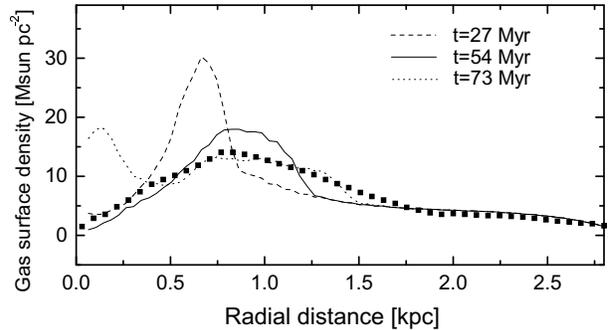}}
      \caption{Azimuthally averaged radial gas distribution obtained with 
      model~1 at three different phases of the shell expansion. The {\it
      measured} radial gas distribution of Ho~I is plotted with the filled 
      squares where a possible contribution of He and $H_2$ to the total 
      gas mass is taken into account (${\rho_{\rm HI+He+H_2}/ 
      \rho_{\rm HI}}=1.4$).}
         \label{Fig8}
\end{figure}

Finally, in Fig.~\ref{Fig8} we present the azimuthally averaged radial gas 
distribution at three different phases of the shell expansion, taking the 
morphological center of the projected shell as the origin. An inclination 
i = 15\degr\, is assumed. The best agreement with observations is found at 
t = 55 $\pm$ 5~Myr, which implies that the shell in Ho~I has attained its 
largest size at the present time. The lack of expansion motions reported 
by Ott et al. (\cite{Ott}) is thus not due to the low inclination of Ho~I, 
but rather an intrinsic property. The measured contrast ($\sim 15$) in gas
surface density between the central depression and the ring is well 
reproduced. We varied the total number of SN explosions ($N_{\rm SN}$)
in order to see how the energy input can influence our conclusions. As 
mentioned above, for $N_{\rm SN} \le 150$ the shell never expands to its 
present diameter of 1.7~kpc, thus ruling out a small starburst as the 
origin of the HI ring in Ho~I. On the other hand, for $N_{\rm SN} \ge 450$ 
the projected shell appears too sharp-peaked at the time when it has reached 
a 1.7~kpc diameter, very similar to the radial gas profile for $N_{\rm SN} =
300$ at 27~Myr (see Fig.~\ref{Fig8}, dashed line). Note that a possible 
depletion of HI column density ($N_{\rm HI}$) due to the ionization by an
external UV background with $\sim 10^6$~$\rm Ly\alpha$ photons 
cm$^{-2}$~s$^{-1}$~Hz$^{-1}$ accounts for $\triangle N_{\rm HI} 
\sim 3\times 10^{19}\: N_{\rm HI}^{-1}$~cm$^{-2}$ only, which is less 
than $14\%$ at $t\ge 54$~Myr. Hence, we find that the energy output 
equivalent to $300\pm50$~SNe can be responsible for producing the HI 
ring-like distribution of Ho~I. If we assume a star formation efficiency
of $5\%-10\%$ and a Salpeter IMF, the mass of a parent gas cloud must be
around $1.5-3.0 \times 10^5~M_{\odot}$, which is not extreme.

\subsection{Slowly rotating galaxy with a small halo (model~2).}
\label{model2}
\begin{figure*}
 \centering
  \includegraphics[width=14 cm]{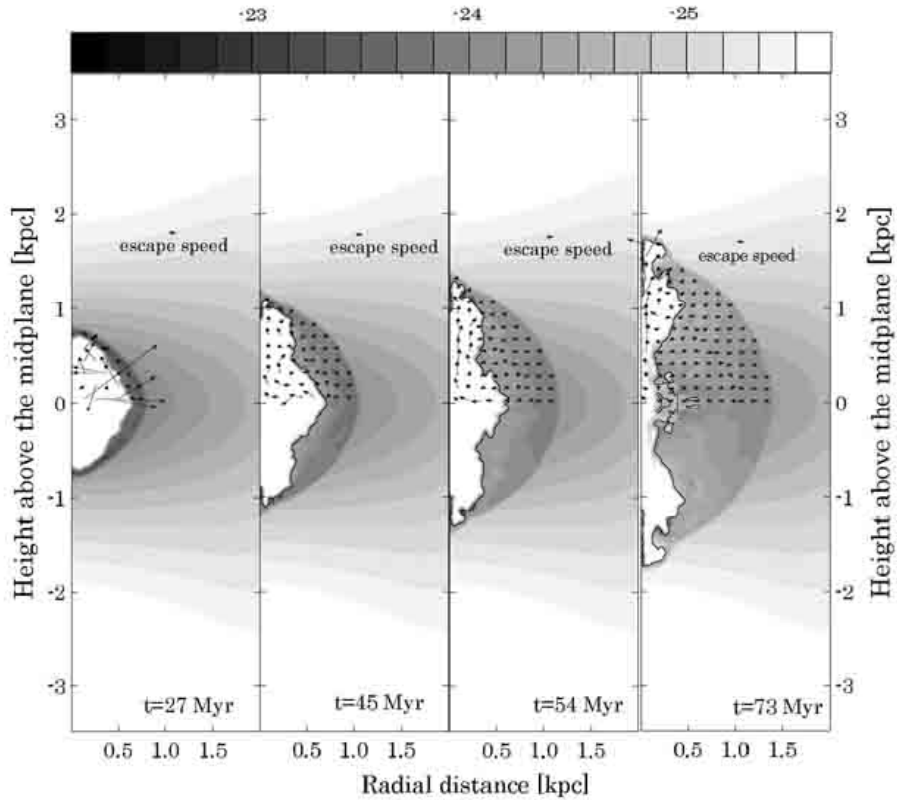}  
      \caption{Time evolution of the gas volume density obtained with 
      model~2, with an energy input equivalent to 150 successive SN 
      explosions. The velocity field is normalized to the local escape 
      speed. The contour line delimits the region filled with SN-ejected 
      material. The grey-scale wedge is in units of g~cm$^{-3}$.}
         \label{Fig9}
\end{figure*}
Our numerical hydrodynamical simulations reveal considerable differences
in the dynamics of the SN-driven shell for the case of a slowly rotating 
galaxy with a small, cuspy halo (model~2), as compared to the fast rotating
galaxy with a massive, shallow halo (model~1). First, we find that the energy 
input needed to create the observed ring-like morphology is a factor of 2
lower in model~2 than in model~1. The radial gas surface density profiles in 
both models are similar; however, the exponential scale height of the gas is 
higher in model~2 (see Fig.~\ref{Fig3}), which results in a smaller gas 
volume density. This fact, along with a much lower inward gravity pull by a 
small halo, renders the isotropic expansion of the shell much more easy, 
but on the other hand impedes break through. Fig.~\ref{Fig9} shows the time
evolution of the distribution of the gas volume density for an energy 
deposition equivalent to 150~SN explosions. The shell evidently never 
breaks out of the disk. This is mainly due to the thicker vertical gas 
distribution and the lower energy input in model~2, as compared to model~1.
On the contrary, the lack of gas expansion in the central HI depression 
in Ho~I reported by Ott et al. (\cite{Ott}) implies that its shell has broken
out of the disk.

Applying the same analysis as in Sect.~\ref{model1}, we find the best 
agreement with the HI spectrum  and channel maps of Ho~I for inclination 
angles i = 25\degr~-~30\degr, in contrast to i = 15\degr~-~20\degr\, for 
model~1. Thus, numerical models with different halo masses predict 
different angles of inclination for Ho~I. Comparison with the inclination 
angle derived from observations could therefore help to constrain the halo 
mass of Ho~I. Unfortunately, Ott et al. (\cite{Ott}) were not able to get 
stable results for the inclination. This implies that Ho~I has a rather 
low inclination, which favours model~1 as the prime choice of Ho~I. 
Furthermore, in case of Ho~I we do not see any HI along the line of sight 
to the central hole down to a column density of $\sim 6.0 \times 
10^{19}$~cm$^{-2}$, or a surface density of $0.45~M_{\odot}$~pc$^{-2}$, 
a feature which is much better reproduced in model~1 
than in model~2. For instance, the gas surface density near the central
depression in Fig.~\ref{Fig10} never falls below $6~M_{\odot}$~pc$^{-2}$.
Indeed, the shell has not yet broken out of the disk in 
model~2, thus contributing substantially to the line-of-sight gas column 
density near the central depression. This tendency is clearly seen in 
Fig.~\ref{Fig10}, which shows the azimuthally averaged radial gas 
distribution at three different phases of the shell expansion, taking 
the morphological center of the projected shell as the origin. An 
inclination of i = 25\degr\, is assumed. As is seen, the surface density
of the gas near the center always exceeds that of Ho~I shown by the filled 
squares. 

\begin{figure}
  \resizebox{\hsize}{!}{\includegraphics{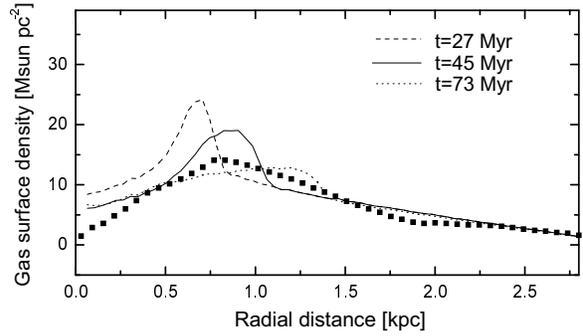}}
      \caption{Azimuthally averaged radial gas distribution obtained with 
      model~2 at three different phases of the shell expansion. The 
      {\it measured} radial gas distribution of Ho~I is plotted with 
      the filled squares. }
         \label{Fig10}
\end{figure}

\section{Discussion}
\label{discus}
Ho~I has a ring-like HI distribution, which is moderately
non-axisymmetric with respect to its morphological center (see Fig.~\ref{Fig1}). 
The ellipticity of the ring makes it rather difficult 
to draw any quantitative comparison with the observations on the basis 
of two-dimensional axisymmetric simulations. However, the basic qualitative 
features of the gas distribution and velocity structure in Ho~I, such as the 
size of the ring, the contrast in the gas surface density 
(hereafter, $\Sigma_{\rm g}$) between the central 
depression and the ring, and the lack of gas expansion in the central hole
can be modelled in the two-dimensional case.

The lack of gas expansion in the central depression  reported
by Ott et al. (\cite{Ott}) argues that the shell has suffered a blowout.
Otherwise, one would expect the shell to be still in the Sedov
expansion phase with some piled-up gas in the z-direction.
However, we do not see any HI along the line of sight 
to the central hole down to a column density of $\sim 6.0 \times 
10^{19}$~cm$^{-2}$. The measured contrast in $\Sigma_{\rm g}$
between the central depression and the ring (a factor of ~ 15) is
also indicative of a blowout, since the `limb brightening' effect
may account for a factor of 2 difference in $\Sigma_{\rm g}$
between the projected rim of a three-dimensional shell and its center.
Hence, the occurrence of a blowout (implied by the measured contrast 
in $\Sigma_{\rm g}$ between the central depression and the ring) at the time when the shell
has expanded out to its present size of 1.7~kpc  
is the main criterion for the selection between the two models.

As discussed in Sect.~\ref{model1} and \ref{model2}, the blowout scenario of Ho~I 
is better reproduced in model~1 at $t=55 \pm 5$~Myr.
At a first glance, the radial gas distribution
in model~2 at 73 Myr seems by eye to better fit the
overall observational data. However, it is not only 
the measured size of the ring (1.7 kpc), but also the measured 
contrast in $\Sigma_{\rm g}$ 
between the central depression and the ring (a factor of ~ 15) 
that are not reproduced for this case. We would like
to note here that the estimated decrease in HI column density near the central
depression caused by the external UV background is approximately 
$2\%$ at $t>45$~Myr in model~2. Hence, the external UV radiation field cannot
reconcile the measured contrast of ~ 15. Although the overall radial
gas distribution in model 2 at 73 Myr fits better the measured profile,
it fails to reproduce both the size of the ring and the measured contrast
in the gas surface density between the central depression and the ring.

On the other hand, our best model is not optimized to fit
the observations. At the time when the ring has attained its present size of
~1.7 kpc, the maximum gas surface density exceeds that of Ho~I by approximately 
$5~M_{\odot}$~pc$^{-2}$ (see Fig.\ref{Fig8}, the solid line). 
At the same time, the width of the modelled ring is smaller 
than that of the observed ring. This is most probably due to 
the two-dimensional axisymmetric nature of our simulations
attempting to reproduce a moderately non-axisymmetric structure, or
azimuthal averaging of the inclined ring with vertical walls. 
Three-dimensional non-axisymmetric modelling of Ho~I is required 
for a fine tuning of the model radial gas distribution.

There are a few assumptions inherent to the model that need further justification: \\
{\it The assumed cooling function}.  The cooling function below $10^{4}$~K 
depends on the degree of ionization of the gas. We made a few test runs with
the cooling function of Spaans \& Norman (\cite{Spaans}) truncated at 
the lower temperatures, i.e. the cooling was 
set to zero at $T<10^4$~K, and found that the influence of the cooling 
at $T<10^4$~K on our results was minimal, which is due to the 
fact that thermal energy of the shell is mostly lost at higher 
temperatures as mentioned earlier.\\
{\it Injection of energy by supernova explosions}. In the present simulations
we use a constant wind approximation (Mac Low \& Ferrara \cite{MF}). 
However, our simulations have shown that 
the injection of thermal energy due to discrete 
SN explosions produce similar results. Indeed, the frequency
of SN explosions in an instantaneously born stellar cluster remains 
nearly constant, if a Salpeter initial mass function
and an upper stellar mass of $100~M_{\odot}$ are assumed. This finding
was also confirmed in numerical simulations of 
Mac Low \& McCray (\cite{MM}). \\
{\it Star formation history}. In the present simulations we have assumed 
a single starburst localized in the central region of Ho I
that provides the energy for 30~Myr.
Constant star formation in Ho~I can be ruled out because of
the low total $H\alpha$ luminosity of $4.3 \times 10^{38}$ ergs s$^{-1}$ 
(Miller \& Hodge \cite{Miller}), which implies a SFR of only $0.004~M_{\odot}$~yr$^{-1}$. 
Indeed, assuming the constant SFR
of $0.004~M_{\odot}~$yr$^{-1}$ and an age of Ho~I of 10~Gyr, one obtains an
estimate on the total stellar mass of $4 \times 10^7 M_{\odot}$, which is
2.5 times lower than that actually measured for Ho~I (Ott et al. \cite{Ott}).
Variable star formation is plausible. However, in dIrr's variations have 
a burst-like nature (see e.g. Searle et al. \cite{Searle}). 
Our model corresponds to an explosive energy input following 
evolution of a single OB association 
with a total mass of $\sim 3 \times 10^4~M_{\odot}$.
A comparison of the photometry presented in Ott et al. (\cite{Ott}) 
with the STARBURST99 synthesis
models (Leitherer et al. \cite{Star}) shows that the B-band
magnitude as well as the U$-$B color of the optical emission within
the giant HI hole is in agreement with a stellar cluster 55~Myr of age
(predicted by model~1) and a mass of $\sim 2\times 10^{5} M_{\odot}$
(Salpeter IMF, Z=0.004 metallicity, upper and lower mass cutoff: $100~M_{\odot}$ and $1~M_{\odot}$, respectively). Such a stellar population
can easily provide the mechanical energy input to Ho~I needed to
satisfy model~1. The discrepancy of the observed stellar mass with our
simulations (about 1 order of magnitude) is most likely due to
confusion with older stellar populations.

\begin{figure}
  \resizebox{\hsize}{!}{\includegraphics{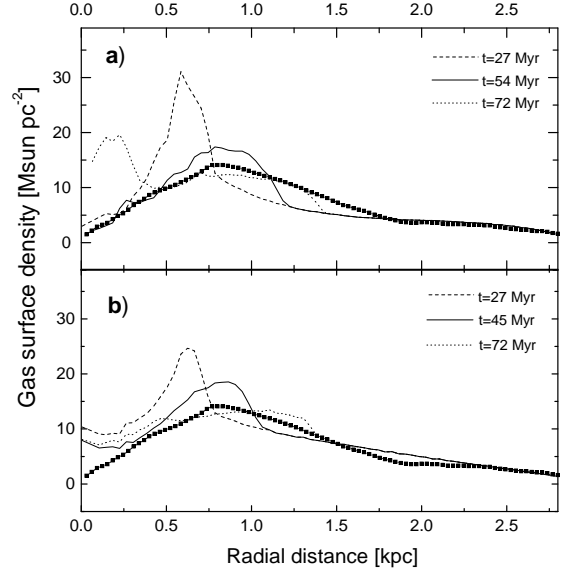}}
      \caption{Azimuthally averaged radial gas distribution obtained 
      for a time-varying mechanical luminosity with 
      {\bf a)} model~1 and {\bf b)} model~2 at three different phases 
      of the shell expansion. The {\it measured} radial gas distribution 
      of Ho~I is plotted with the filled squares. }
         \label{Fig11}
\end{figure}

Nevertheless, we examine if variable star formation can affect our conclusions. 
We assume that the mechanical luminosity produced by SN explosions 
is time-dependent, for instance with a Gaussian shape. 
Its maximum is centered at 15~Myr after the beginning of the energy input phase, 
with a FWHM of 8.5~Myr. 
This time-dependent mechanical luminosity is supposed to mimic 
the energy input from successively born stellar clusters within a single stellar
association. To be consistent with the simulations in Sect.~\ref{results},
we set the energy input phase to last for 30~Myr and the total energy input
equivalent to 300 and 150 SNe for model~1 and model~2, respectively.
Figure~\ref{Fig11} shows the azimuthally averaged gas distributions obtained
at the same phases of the shell expansion as in Figs.~\ref{Fig8} and \ref{Fig10}.
As is obviously seen, model~1 with variable star formation reproduces 
the observed distribution and the contrast in $\Sigma_{\rm g}$
between the central depression and the ring (the solid line, Fig.~\ref{Fig11}a) 
even better than
the same model with a single starburst approximation (the solid line, Fig.~\ref{Fig8}).
Model~2 clearly fails to reproduce the measured contrast in $\Sigma_{\rm g}$ for
both star formation frameworks.
Our simulations also show that multiple, successively born smaller clusters 
are less effective in creating a bubble structure than a single starburst 
with the same total ejected energy. \\
{\it Numerical resolution}.  
We made a few test runs with a higher resolution of 5 pc. Although the higher
resolution simulations provide more details on the gas flow dynamics and
the development of Rayleigh-Taylor and Kelvin-Helmholtz instabilities 
(Figs.\ref{Fig4} and \ref{Fig9}), the integrated images such as the HI spectrum, 
the channel maps, and 
the radial gas distributions (Figs.~\ref{Fig8}, \ref{Fig10}, and \ref{Fig11}) 
remain very similar to those derived with a lower resolution of 10 pc.

\section{Summary}
\label{concl}
Numerical hydrodynamical modelling of the SN-driven shell formation is
performed with the aim to reproduce the ring-like morphology of HI in
the dIrr Ho~I. Two models of Ho~I are considered: a fast rotating galaxy 
with a massive halo of $6.0\times 10^9~M_{\odot}$ (model~1), and a slowly 
rotating galaxy with a small halo of $4.0\times 10^8~M_{\odot}$ (model~2). 
Simulations reveal the following differences between the two models: \\
\indent
1. The mechanical energy to create the ring of 1.7~kpc
diameter in model~1, $(3.0 \pm 0.5) \times 10^{53}$~ergs or $300 \pm 50$ 
Type~II SNe, is a factor of 2 higher than in model~2. This is mainly due 
to the smaller gas volume density and weaker gravitational potential in 
model~2 as compared to model~1. 

2. In model~1 the shell breaks out of the disk before it expands out 
to its present size of 1.7~kpc. On the contrary, in model~2 the shell 
never breaks out of the disk, mostly due to the thicker vertical gas 
distribution and lower energy input. As a consequence, both the measured 
contrast ($\sim 15$) in gas surface density between the central depression 
as well as the ring and the lack of gas expansion in the center of 
the ring are better reproduced in a model with a massive ($6.0 \times 
10^9~M_{\odot}$) dark matter halo. 

3. Within model 1 an upper (i = 
20\degr) and lower (i = 15\degr) limit to the inclination of Ho~I can 
be set from the modelled HI spectrum and the channel maps, respectively.
This is probably the only handle to assess the inclination of gas-rich,
nearly face-on dwarf galaxies.

Our numerical hydrodynamical simulations suggest that Ho~I was initially 
a halo-dominated dIrr with an exponentially declining radial gas 
distribution. Approximately 60~Myr ago a stellar cluster was born near 
the dynamical center of the galaxy from a parent gas cloud of $(1.5 - 
3.0)\times 10^5~M_{\odot}$. Subsequent SN explosions created an expanding 
shell filled with the hot SN ejecta. About 15~Myr ago the shell broke out 
of the disk pumping the hot gas into the halo. The fraction of the total 
gas mass of Ho~I lost in a blow-out is low, within $1-2\%$. 
At present the shell has attained its biggest 
size of 1.7~kpc. In another 20~Myr the central hole would fill in and Ho~I 
would appear as a dIrr, with a wave-like declining radial gas distribution.

\begin{acknowledgements}
The authors are thankful to M. Spaans and C. Norman for providing 
their cooling curves. This work was done under the INTAS grant YSF-2002-33. 
EV is grateful to the staff of the Radioastronomisches Institut der 
Universit\"at Bonn for their hospitality and help with data processing.  
YS acknowledges partial support from the German Science Foundation,
DFG (project SFB 591, TP A6). We are thankful to the anonymous referee for valuable criticism. 
\end{acknowledgements}

\end{document}